\documentclass[preprint]{article}
\usepackage{graphicx}% Include Figure files
\usepackage{dcolumn}% Align table columns on decimal point
\usepackage{bm}% bold math
\usepackage{amsmath}
\usepackage{amssymb}
\usepackage{multirow}
\usepackage{caption}
\usepackage{epstopdf}
\usepackage{hyperref}
\begin{document}
{\setlength{\oddsidemargin}{1.2in}
\setlength{\evensidemargin}{1.2in} } \baselineskip 0.55cm
\begin{center}
{\LARGE {\bf Gravastar model in the structure of $f(R,L_{m}, T)$ modified theory of gravity }}
\end{center}
\date{\today}
\begin{center}
  Meghanil Sinha*, S. Surendra Singh \\
Department of Mathematics, National Institute of Technology Manipur,\\
Imphal-795004,India\\
Email:{ meghanil1729@gmail.com, ssuren.mu@gmail.com}\\
 \end{center}

\textbf{Abstract}: The Gravastar ( or the Gravitational Vacuum Star ) is a very serious alternatives proposed to the principle of the Black Hole, the model of which was originally developed by Mazur and Mottola. A Gravastar is an astronomically hypothetically condensed object which is a gravitationally dark vacuum star or a gravitational vacuum condensate star, which is singularity free, spherically symmetric and also super compact. The current study concerns about the model of the Gravastar in the modified $f(R,L_{m}, T)$ gravity considering the form $f(R,L_{m},T) = R +  {\alpha}TL_{m} $. From Mazur-Mottola \cite{Mazur}-\cite{Mottola}, we get to know that a Gravastar model has three distinct regions having various Equations of State (EoS). We have inquired into the interior portion with the space-time considering $ \rho = - p $, for the dark sector of the interior region, here the negative matter-energy density exerting a repulsive force on the immediate thin shell with the EoS $ \rho = p $ where it is considered as an ultra-relativistic fluid. We have studied the properties such as energy density, proper length, total energy and entropy. Next comes the vacuum exterior region of the Gravastar which is being described by the Schwarzschild-de-Sitter solution. And also from Darmois-Israel formalism, we have probed the junction connecting the inner and the outer surfaces of the Gravastar. \\


 \section{Introduction}\label{sec1}

The first ever proposal to gravitationally vacuum condensate star was done by Mazur-Mottola as a solution obtaining from the idea of Bose - Einstein condensation applied to gravitational systems \cite{Mazur}-\cite{Mottola}. The framework of Gravastar was proposed as a singularity-free condensed object, which served as a substitution of a Black Hole with no event horizons. Further it was explored as the compact stars which don't have event horizons. 
Black Holes, regarded as a fascinating topic in General Relativity, the major reasons being, they have entropy, arising solely from quantum mechanical effects \cite{Hawking}. Even a collapse in non-spherical symmetry could also lead to Black Hole singularity shown by  \cite{Penrose}, being the exact solutions to the Einstein's equations.  Black Holes or massive compact objects was observationally found by \cite{Ghez, Gillessen}.
 The Black Hole is being considered as an ultimate fate of a star depending on the initial mass of the collapsing stellar object. The first to get the solution for the Black Hole was by Karl Schwarzschild in the theory of General Relativity. From the starting point of the solution for a Black Hole, it suffered from two major problems, the problem of singularity at the center and the event horizon(point of no return) problem. The event horizon which existed at $ r = 2M $ ($ M $ = mass of the Black Hole, all are in astronomical units) creates a boundary on Black Hole and exterior space-time, below which all physical laws become irrelevant, here due to the infinite curvature of space-time and moreover the gravitational pull is so severe, strong that our current understanding of physics breaks down. Thus singularity and the event horizon which exist at the final state of stellar collapse, that is Black Hole, have always led to a problem in the field of research. These two are the major drawbacks in the success of the Black Hole theory. 
  For a compact overview of the Black Holes, an alternate model was required, to overcome the central singularity and also the event horizon problem. According to Mazur and Mottola, phase transition avoids further collapse of a star and thus proposed the gravitational vacuum star model or rather the Gravastar model in this context. Gravastars look similar to Black Holes but they have no singularity and no event horizon. These types of stellar objects help to describe the role of dark energy in the accelerated expansion of the Universe and also help to explain why some particular Galaxies have low or high concentration of dark matter. \\
  The idea of Gravastar is originated owing to the fact that during the phase transition, an entire condensed matter system undergoes a phase transition. Due to non-negligible quantum mechanics near the event horizon, the behaving of the collapsing dust particle like a quantum system with multiple interactions can be expected. We know that a group of Bosonic atoms or molecules go to Bose-Einstein condensation in a very low temperature. Mazur and Mottola studied the Bose-Einstein condensation and extended it in celestial bodies in gravitational collapse where they build in the idea of constructing a hypothetical compact cold and dark object, and named it as gravitationally vacuum condensate star or Gravastar to overcome the problems of singularity and event horizons. Thus the model was considered as a different approach to the classical Black Hole. Many researchers have explored the Bose-Einstein condensate(BEC) in Astrophysics, and Boson Stars \cite{Panotopoulo} is expected to be formed from BEC and the core of Neutron Stars is from BEC. The originally five-layered Gravastar model as proposed by Mazur and Mottola was reduced to a three-layered final model of the Gravastar by \cite{Visser Wiltshire}. The interior region with the de-Sitter condensate or anti-de Sitter phase with the EoS  $ \rho = - p $, the intermediate thin shell having EoS $ \rho = p $ and the exterior completely vacuum described by the Schwarzschild manifold. The centre of this Gravastar model is known as dark energy, and the boundary of the shell isolating its interior and exterior regions . The thin shell replaces the conception of the event horizon of the classical Black Hole feature.\\
 After the theory of Gravastar model gained popularity in the field of research, a lot of researchers have been involved in the study of Gravastars in different modified gravitational theories. Gravastar model in modified theories of gravity like $ f(G,T) $ \cite{Gravity2} , $ f(Q,T) $ \cite{Gravity3} , $ f(Q) $ \cite{Gravity4} , $ f (T) $ \cite{Gravity5} , $ f(R,T^2) $ \cite{Gravity6} , $ f(R,G) $ \cite{Gravity8} can be found. Further researches gained insight into charged Gravastar model with exterior Schwarzschild geometry being replaced by Reissner–Nordström solution in the following modified alternate and extended gravity theories$ f(R,T) $ \cite{Charge1} , $ f(Q) $ \cite{Charge2} \cite{Charge3}, Rainbow - Rastall gravity \cite{Charge5}. Different Gravastar solutions in cylindrically space-time can be found in references \cite{Cylinder2} \cite{Cylinder3}. Charged Gravastar models having conformal motion is studied in $ f(T) $ gravity \cite{Conformal1} and also in higher dimensional space-times. Charged Gravastar models with conformal motion is also studied in \cite{Conformal2} \cite{Conformal3}. Charged Gravastars consisting of conformal motion iis investigated by \cite{Conformal4} . Further researches on Gravastar model in  $ (2+1) $ dimensional anti de-Sitter space-time can be identified in\cite{Dimension1} \cite{Dimension2} \cite{Dimension3}. Different research papers and articles on the formation of these gravitational vacuum condensate star models from the Black Holes can be encountered in references \cite{BH1} \cite{BH2}. Gravastar solutions having continuous pressure has been examined in \cite{Prop1}. While Cattoen, Faber, Visser showed that Gravastars should have anisotropic pressures \cite{Prop2}. \\ 
In this paper, we have delved into stable Gravastar model for a particular matter Lagrangian in the foundation of $ f(R,L_{m}, T) $ gravity in the static, spherically symmetric space-time. This manuscript is organized as: In Sec.\ref{sec2} , we have established the formulation of the field equation in the $ f(R,L_{m}, T) $ gravity. In Sec.\ref{sec3} , we explored the modified field equations with spherically symmetric space-time for the Gravastar model with the assumed Lagrangian and the non-conservation equation of the energy momentum tensor in  $ f(R,L_{m}, T) $ theory of gravity. Sec.\ref{sec4} deals with the geometry concerning the Gravastar. Different physical properties such as proper length, energy content, entropy and Equation of State are discussed in Sec.\ref{sec5} , while the junction conditions linking the interior region and the exterior vacuum region of the Gravastar are discussed in Sec.\ref{sec6} with EoS. The concluding remarks and conclusions are provided in Sec.\ref{sec7}. \\  

 \section{Mathematical formalism of the gravitational field equation in  $ f(R,L_{m}, T) $  gravity framework}\label{sec2}

Using the generalization of  $ f(R,T) $ \cite{Lobo} and  $ f(R,L_{m})$ \cite{Harko} space-time descriptions together, the generalized model of the  $ f(R,L_{m}, T) $ gravity model was put forward by \cite{Haghani}, where $ R $ represents an arbitrary function of the Ricci scalar, $ T $ represents the trace of the energy - momentum tensor and $ L_{m} $ the matter Lagrangian thus $ L_{grav} = f(R,L_{m}, T) $ .\\
The Einstein-Hilbert action describing the $ f(R,L_{m},T) $ gravity theory with strong geometry-matter coupling is given by \\
\begin{equation}\label{1}
 S=\frac{1}{16\pi }\int f\left( R,L_{m},T\right) \sqrt{-g}\;d^{4}x+\int {L_{m}%
\sqrt{-g}\;d^{4}x} 
 \end{equation} \\
 where $ g = det(g_{\mu \nu} ) $, $ g_{\mu \nu} $ being the metric tensor and assuming $ c = G = 1 $ throughout this paper. The energy-momentum tensor is defined by
\begin{equation}\label{2}
T_{\mu \nu }=-\frac{2}{\sqrt{-g}}\frac{\delta \left( \sqrt{-g}L_{m}\right) }{%
\delta g^{\mu \nu }}.
\end{equation} \\ 
 Now, varying the action of equation(\ref{1}) with respect to the metric tensor, we get the field equation as \\
 \begin{align}\label{3}
 f_{R}R_{\mu \nu }-\frac{1}{2}\left[ f-(f_{L}+2f_{T})L_{m}\right] g_{\mu
\nu }+\left( g_{\mu \nu }\Box -\nabla _{\mu }\nabla _{\nu }\right) f_{R}=\left[
8\pi +\frac{1}{2}(f_{L}+2f_{T})\right] T_{\mu \nu }+f_{T}\tau_{\mu\nu}.
\end{align}% \\
where $\Box \equiv \partial_{\mu}(\sqrt{-g}g^{\mu\nu}\partial_{\nu})/\sqrt{-g}$, $f_{R}\equiv \partial f/ \partial R$, $f_{T}\equiv \partial f/ \partial T$, $f_{L} \equiv \partial f/ \partial L_{m}$, $R_{\mu\nu}$ is the Ricci tensor,  $\nabla_{\mu}$ the covariant derivative with respect to the symmetric connection associated to $g_{\mu\nu}$, and the new tensor $\tau_{\mu\nu}$ is defined as \\
\begin{equation}\label{4}
\tau_{\mu\nu} = 2g^{\alpha\beta} \frac{\partial^{2}L_m}{\partial g^{\mu\nu} \partial g^{\alpha \beta}}.
\end{equation}\\
Evidently, if $ f(R,L_{m}, T) $ = $ f(R) $, in equation(\ref{3}) , we get the field equations of  $ f(R) $ gravity, if $ f(R,L_{m}, T) = f(R,T) $, we get the field equations for $ f(R,T) $ gravity and when  $ f(R,L_{m}, T) = f(R,L_{m}) $, we get the field equations for the  $ f(R,L_{m}) $ theory. In the most general case, for $ f(R,L_{m}, T) = R $ , we get the standard field equation for The General Relativity, that is 
 $ R_{\mu \nu} - \frac{1}{2}g_{\mu \nu}R = 8\pi T_{\mu \nu} $ . \\
 Taking the covariant divergence of the field equation, equation(\ref{3}), we get  the non-conservation of the energy momentum tensor as
\begin{align}\label{5}
    \nabla^{\mu}T_{\mu\nu} =&\ \frac{1}{8\pi + f_m}\Big[ \nabla_{\nu} (L_m f_m) - T_{\mu\nu} \nabla^{\mu} f_m  \nonumber  \\
    &\left.- A_\nu - \frac{1}{2}(f_T \nabla_\nu T + f_L \nabla_\nu L_m) \right] ,  
\end{align}\\
where we have used the fact that $\nabla^\mu R_{\mu\nu}= \nabla_\nu R/2$ and the mathematical property $(\square\nabla_\nu - \nabla_\nu\square)\phi = R_{\mu\nu}\nabla^\mu\phi$, valid for any scalar field $\phi$ and $ f_m $ and $ A_{\nu} $ is defined as\\
\begin{equation}\label{6}
 f_m = f_T + \frac{1}{2}f_L 
\end{equation}\\
and  
\begin{equation}\label{7}
 A_\nu = \nabla^\mu(f_T \tau_{\mu\nu}).
 \end{equation}
Assuming a perfect fluid coupled to a scalar field, we have $ A_{\nu} = 0 $ and noting that 
\begin{equation}\label{8}
\nabla _{\nu }f\left( R,L_{m},T\right) =f_{R}\nabla _{\nu }R+f_{T}\nabla
_{\nu }T+f_{L}\nabla _{\nu }L_{m},
\end{equation}%\\
and $ \nabla^{\mu}G_{\mu \nu} = 0 $ and using the identity 
\begin{equation}\label{9}
\left( \Box \nabla _{\nu }-\nabla _{\nu }\Box \right) f_{R}=R_{\mu \nu
}\nabla ^{\mu }f_{R},
\end{equation}%
to obtain the above eqation(\ref{5}). Equation(\ref{5}) is the direct consequence indicating the existing matter fields in the expression of the gravitational Lagrange density, given by the functional form of $ f(R,L_{m}, T) $. Clearly, for the case of $ f_T(R,L_{m}, T) = f_L(R,L_{m}, T ) =0 $, the matter content comprising the Universe in conserved. In this paper, we consider the cosmic matter described by a perfect fluid approximation characterized by only two thermodynamics parameters, the energy density $ \rho $ and the thermodynamics pressure $ p $ of the fluid respectively. In this context, the matter energy momentum tensor is given by \\
\begin{eqnarray}\label{10}
T_{\mu\nu}=(p + \rho)u_{\mu}u_{\nu} + pg_{\mu\nu},
\label{momentum_tensor}
\end{eqnarray} \\
where $ u_{\mu} $ is the four - velocity vector satisfying $u_{\mu}u^{\mu}= -1$ .\\
Our motivation with regard to functional representation is $ f(R,L_{m}, T) = R + \alpha T L_{m} $ with $\alpha $ being matter geometry coupling constant. For this model, we get equation (\ref{3}) and equation (\ref{5}) as\\
\begin{equation}\label{11}
G_{\mu\nu}= \left[ 8 \pi + \frac{\alpha}{2}(T + 2 L_m) \right] T_{\mu\nu} + \alpha L_m(\tau_{\mu\nu}- L_mg_{\mu\nu}) 
\end{equation}\\
and\\
\begin{eqnarray}\label{12}
\nabla^{\mu}T_{\mu\nu}= \frac{\alpha}{8 \pi +\alpha (L_m + T/2 )}\left[ \nabla_{\nu}\Big(L_m^{2} + \frac{1}{2}T L_m \Big) - T_{\mu\nu} \nabla^{\mu} \Big(L_m + \frac{T}{2} \Big) - \nabla^\mu(L_m\tau_{\mu\nu}) - \frac{1}{2}(L_m \nabla_{\nu}T + T \nabla_{\nu}L_m) \right] 
\end{eqnarray}\\
respectively, where $G_{\mu\nu}$ is the Einstein tensor. For  $ \alpha = 0 $, we get $\nabla^{\mu}T_{\mu\nu}= 0$ and $G_{\mu\nu}= 8\pi T_{\mu\nu}$, that is the conservation equation in the General relativity theory. 

 \section{Modified field theory equations and their resolvents in $ f(R,L_{m}, T) $ gravity theory}\label{sec3}
 
 We consider here the spherically symmetric metric with the line element given by 
\begin{equation}\label{13}
ds^2=e^{a(r)}dt^2-e^{b(r)}dr^2-r^2(d\theta^2+\sin^2\theta
d\phi^2)
\end{equation}\\
From equation (\ref{4}), clearly $ \tau_{\mu\nu} $ depends on the matter Lagrangian density $ L_{m} $. Two potential outcomes exist for the matter Lagrangian resulting in the possibility of the energy momentum tensor of a perfect fluid (\ref{10}) , that is $ L_{m} = p $ and $ L_{m} = -\rho $ . Here, we are assuming $ L_{m} = - \rho $ , for which we have equation (\ref{11}) as
\begin{equation}\label{14}
G_{\mu\nu}= \left[ 8 \pi + \frac{3 \alpha}{2}(p - \rho) \right] T_{\mu\nu} - \alpha \rho^{2} g_{\mu\nu} 
\end{equation}\\
The non-zero components of the Einstein tensor are given as \\
\begin{equation}\label{15}
G_0^{0}=\frac{e^{-b}}{r^{2}}(-1+e^{b}+b'
r)
\end{equation}

\begin{equation}\label{16}
G_1^{1}=\frac{e^{-b}}{r^{2}}(-1+e^{b}-a'
r)
\end{equation}

\begin{equation}\label{17}
G_2^{2}=G_3^{3}=\frac{e^{-b}}{4r}[2(b'-a')-(2a''+a'^{2}-a'b')r]
\end{equation}
where prime indicates the derivative with respect to the radial
coordinate $r$. Substituting the above, the field equations can be formulated as \\
\begin{equation}\label{18}
\frac{e^{-b}}{r^{2}}(-1+e^{b}+b'r) =  8 \pi \rho + \frac{3 \alpha}{2}(p - \rho) \rho - \alpha \rho^{2}
\end{equation}

\begin{equation}\label{19}
\frac{e^{-b}}{r^{2}}(-1+e^{b}-a'r) = - 8 \pi p + \frac{3 \alpha}{2}(p - \rho) p - \alpha p^{2}
\end{equation}

\begin{equation}\label{20}
\frac{e^{-b}}{4r}[2(b'-a')-(2a''+a'^{2}-a'b')r] = - 8 \pi p + \frac{3 \alpha}{2}(p - \rho) p - \alpha p^{2}
\end{equation}\\
Here from the equation(\ref{18}), we can have 
\begin{equation}\label{21}
e^{-b}=1-\frac{8\pi \rho r^2}{3}-\frac{\alpha}{2}(3p-5\rho)\frac{\rho r^2}{3}.
\end{equation}\\
or we have it as \\
\begin{equation}\label{22}
e^{-b}=1-\frac{2m}{r}-\frac{\alpha}{2}(3p-5\rho)\frac{\rho r^2}{3}
\end{equation}\\
by considering $ m $ as the gravitational mass enclosed within the sphere of radius $ r $.
 We get from the non-conservation equation of the energy-momentum tensor equation (\ref{12}) , \\
\begin{equation}\label{23}
    p' + \frac{a'}{2}(\rho + p) + \frac{\alpha\left[ 4\rho\rho'+ 3p(\rho'- p') \right]}{16\pi + 3\alpha (p - \rho)} = 0 .
\end{equation}\\
From the equations (\ref{18}) - (\ref{20}) with equations (\ref{21}) - (\ref{22}) and equation  (\ref{23}), we get the hydrostatic equilibrium equation for the stellar system with spherically static symmetrical structure  in  $f(R, L_m,T)= R+ \alpha TL_m$ model having $L_m = -\rho $ , \\
\begin{align}\label{24}
    \frac{dp}{dr} &= -\frac{(\rho+ p) \left[4\pi r p +\frac{m}{r^2} + \frac{3\alpha r}{4}(p-\rho)p - \frac{\alpha r}{2}p^2 + \frac{\alpha r}{12}(3p-5\rho)\rho \right]}{\Big(1 - \frac{2m}{r} -\frac{\alpha}{2}(3p-5\rho)\frac{\rho r^2}{3}\Big)\left\{1 + \frac{\alpha\left[3p(d\rho/dp-1) + 4\rho (d\rho/dp) \right]}{16\pi + 3\alpha(p - \rho)} \right\}} 
\end{align}\\
Using the energy density $ \rho $ dependence criterion on the pressure $ p $, a barotropic EoS that is $p= p(\rho)$, so that $\rho' = (d\rho/dp)p'$. For $ \alpha = 0 $ , we can reduce it to the standard form of the Tolman - Oppenheimer - Volkoff (TOV) equations as in the case of the theory of General Relativity .

 \section{Geometry of gravitational vacuum condensate stars}\label{sec4}

We explore here the separate regions of the Gravastar's structure, mainly the Interior Region, then the Intermediate thin shell and then outer space-time geometry. \\
\subsection{Interior space-time}

 Following Mazur - Mottola's \cite{Mazur} \cite{Mottola} approach, assuming the EoS for the interior region of the Gravastar as \\
\begin{equation}\label{25}
p = -\rho 
\end{equation}
The equation of state(EoS) is of the form $ p = w \rho $ , with $ w $ as the Equation of State parameter with $ w = -1 $ here, known as the dark energy equation of state. Using this and the non-conservation of the energy momentum tensor, we can have 
 \begin{equation}\label{26}
\rho = \rho_{k}(constant) 
\end{equation}
and thus the pressure becomes 
\begin{equation}\label{27}
p = - \rho_{k} 
\end{equation}
Substituting $ p $ and $ \rho $ in the equation (\ref{18}), we obtain,
\begin{equation}\label{28}
e^{-b}=1-\frac{4 \rho_k r^2}{3}(2\pi -\alpha \rho_k) + \frac{A}{r}
\end{equation}\\
where $ A $ being the integration constant. From the singularity condition, that is since the Gravastar model is singularity free, thus assuming regular at $ r = 0 $ , we can set $ A = 0 $. Hence, we have 
\begin{equation}\label{29}
e^{-b}=1-\frac{4 \rho_k r^2}{3}(2\pi -\alpha \rho_k) .
\end{equation}
Using the equations (\ref{26}) and (\ref{27}) in the field equations (\ref{18}) and (\ref{19}), we get the interconnections between the space-time potentials as  
\begin{equation}\label{30}
e^{a} = He^{-b} 
\end{equation}
where $H$ being the integration constant.
\begin{figure}
  \centering
  \includegraphics[scale=0.5]{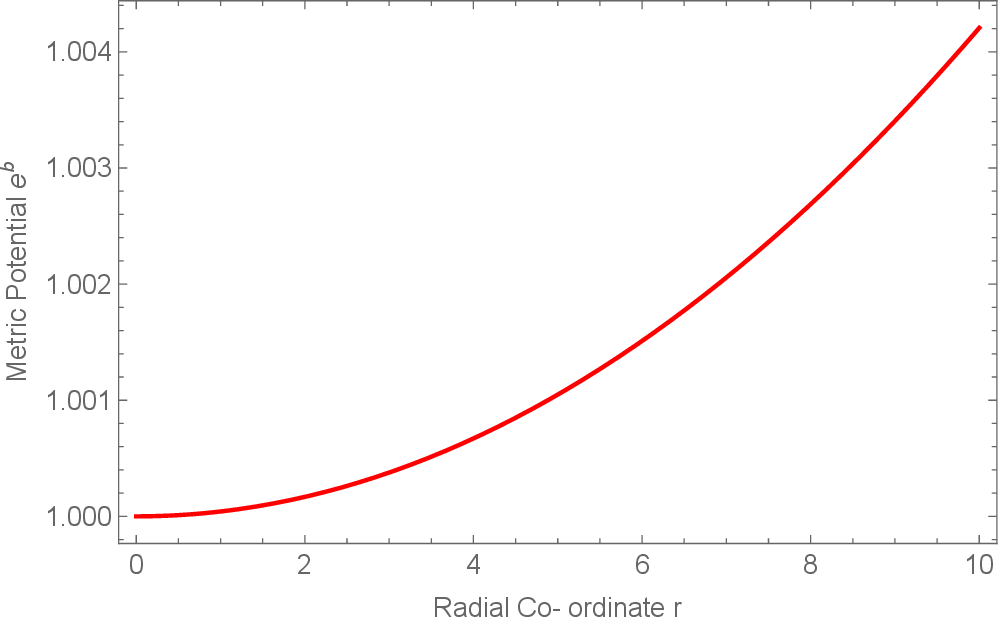}
  \caption{Figure representing the radial dependence of the metric coefficient $ e^{b} $ in the interior of the Gravastar }\label{1}
\end{figure}
The space-time of the Gravastar's interior is singularity free at the center. We get the central active gravitational mass as 
\begin{equation}\label{31}
M(D)= \int_0^{R_1=D} 4\pi r^2{\rho_k} dr=\frac{4}{3}\pi
D^3\rho_k
\end{equation}
where we have considered $ R_1 $ as the boundary of the interior region of the Gravastar with $ r $ as the radial co-ordinate. \\
Figure \eqref{1} represents the variation of the metric potential $ e^{b} $ with respect to the radial co-ordinate $ r $. Here, clearly the metric function stays positive in the interior region and clearly regular at $ r = 0 $ having no central singularity. 

\begin{figure}
  \centering
  \includegraphics[scale=0.5]{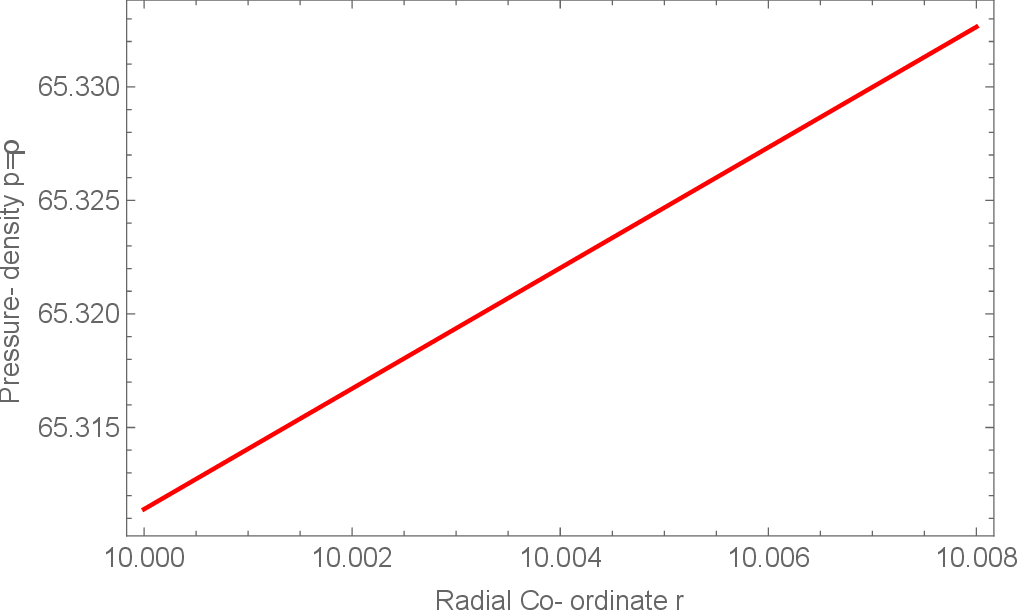}
  \caption{ Graphical representation of pressure $ p $ of the ultrarelativistic fluid inside the thin shell in terms of radial position $ r(km) $ }\label{2}
\end{figure}

\subsection{Shell}

We have taken into consideration that the shell to be composed of highly relativistic fluid with equation of state 
\begin{equation}\label{32}
p = \rho 
\end{equation}
The concept of stiff fluid coupled to the cold baryonic matter is proposed in \cite{Zeldovich}. In the shell region, for the purpose of calculation, we are assuming $0< e^{-b}\ll1$ in the ultra-relativistic thin shell, where two space-times are joining together. In the shell, $r\rightarrow 0$ indicates that any parameter depending on the radial co-ordinate is  $\ll1$ . With these assumptions, we get the field equations  (\ref{18}) - (\ref{20}) with the substitution $ p = \rho $ as
\begin{equation}\label{33}
\frac{e^{-b} b'}{r} + \frac{1}{r^2} = 8\pi \rho - \alpha \rho^2 
\end{equation}
\begin{equation}\label{34}
\frac{1}{r^2} = -8\pi p + \alpha \rho^2 
\end{equation}
\begin{equation}\label{35}
(\frac{3}{2r} + \frac{a'}{4})(e^{-b}.b') = -8\pi p + \alpha \rho^2 
\end{equation}\\
Solving which, we get
\begin{equation}\label{36}
e^{-b}  =  2\ln r + N
\end{equation}
N being the integration constant and with  $r$ being 
$D\leq\,r\leq{D+\epsilon}$. Subject to the constraint $\epsilon\ll1$ , we get $N\ll1$ as $e^{-b}\ll1$ and also we get ,
\begin{equation}\label{37}
e^{a}=Gr^{-4}
\end{equation}
where $G$ is an integration constant.Substituting the relation $ p = \rho $ in the energy equation (\ref{23}) we get $ p(r) = \rho(r) = \frac{4\pi W(\frac{r^{4}\alpha}{4\pi})}{\alpha} $ where  $ W(z) $ denotes the product logarithm function or the Lambert function. Figure \eqref{2} represents how the nature of the pressure $ p $ changes with respect to the radial co-ordinate $ r $, in the shell which is being filled by highly relativistic matter.

\subsection{Exterior space-time}

The outside of the Gravastar is described as vacuum with the static exterior Schwarzschild solution and with the equation of state 
\begin{equation}\label{38}
p = \rho = 0 
\end{equation}
is 
\begin{eqnarray}\label{39}
ds^2=\left(1-\frac{2M}{r}\right)dt^2-\left(1-\frac{2M}{r}\right)^{-1}dr^2 -r^2\left(d\theta^2+sin^2\theta
d\phi^2\right)
\end{eqnarray}
where $M$ being the mass of the gravitational system .

\section{Physical attributes of the Gravastar model}\label{sec5}

Here we have traversed different physical characteristics of a Gravastar model in the realm of $ f(R,L_{m}, T) $  gravity such as proper length or thickness of the shell, shell's energy content, the measure of disorderness, that is entropy content inside the shell.\\

 \subsection{Proper length of the Shell}
 
 The perfect fluid with high stiffness that generates between the radius of the inner boundary of the shell of the Gravastar, i.e.  $ r= D $ and radius of the outer perimeter of the shell $ r = D + \epsilon $ , where $ \epsilon $, which indicates the proper shell thickness ( i.e., $\epsilon\ll1$ ) is assumed to be very small. The proper length of the shell is given by \\
 \begin{equation}\label{40}
\ell= \int_D^{D+\epsilon}
\sqrt{e^{b}}dr=\int_D^{D+\epsilon}\frac{dr}{\sqrt{2\ln r +
N}}.
\end{equation}\\
Integrating the above, we get 
\begin{equation}\label{41}
\ell=\left[
\left(\frac{\pi}{2}\right)e^{\frac{-N}{2}}~erfi\left\lbrace{\sqrt{\left(\frac{N}{2}\right)+\ln r}}\right\rbrace\right]_D^{D+\epsilon}
\end{equation}\\
where $ erfi(x) $ denotes the imaginary error function .\\

\subsection{Energy content}

The interior of the Gravastar with the EoS $ p = -\rho $ forming the region of negative or dark energy density which indicates the repulsive nature existing in the interior of the Gravastar. Thus the energy content within the shell is thus given by 
\begin{eqnarray}\label{42}
\mathcal{E}=\int_{D}^{D+\epsilon}4\pi\rho\,r^{2}dr
\end{eqnarray}\\
where 
\begin{equation}\label{43}
 p(r) = \rho(r) = \frac{4\pi W(\frac{r^{4}\alpha}{4\pi})}{\alpha}
\end{equation}

\subsection{Entropy within the Shell}

Entropy means disorderness within the Gravastar scenario. Mazur-Mottola \cite{Mazur} \cite{Mottola}, proposed the entropy free in the interior space-time with a single-phase condensate, whereas the entropy which is inside the shell can be measured as 
\begin{equation}\label{44}
S=\int_{D}^{D+\epsilon}4\pi\,r^{2}s(r)\sqrt{e^{b}}dr
\end{equation} 
where particularly $ s(r) $ denotes the local entropy density at a given temperature $ T(r) $, and 
\begin{equation}\label{45}
s(r)=\frac{\gamma^2k_B^2T(r)}{4\pi\hbar^2 } =
\gamma\left(\frac{k_B}{\hbar}\right)\sqrt{\frac{p}{2 \pi
}}
\end{equation}
$ \gamma $ being a dimensionless parameter .\\
Thus the entropy can be framed as 
\begin{equation}\label{46}
S(r)
={\frac{4\pi\alpha}{\sqrt{2\pi}}}\int_{D}^{D+\epsilon}\frac{r^2 {\sqrt{p}}}{\sqrt{2\ln
r + N}}dr = 2{\sqrt{2}}\alpha{\sqrt{\pi}}\int_{D}^{D+\epsilon}\frac{r^2 {\sqrt{p}}}{\sqrt{2\ln
r + N}}dr
\end{equation}\\
where we have 
$ p(r) = \rho(r) = \frac{4\pi W(\frac{r^{4}\alpha}{4\pi})}{\alpha} $ , and where we have assumed $ G = c = 1 $ with the Planckian units also  $ k_B = \hbar = 1 $.  

\section{ Junction conditions between the interior and the exterior regions and equation of state }\label{sec6}

Gravastar being made up of three regions, the intermediate thin shell acts as the junction area for the matching between the interior and the exterior of the Gravastar. Darmois - Israel formalism \cite{Israel} \cite{Darmois} suggested that the matching between the interior and the exterior region of the Gravastar must be smooth enough. The metric coefficients although whose derivatives might not be continuous but there is no  discontinuity at the junction surface ($\Sigma$), i.e., at $r=D$ . \\
The stress energy tensor in this context is expressed at the junction with Lanczos equation \cite{Lanczos} \cite{Sen} \cite{Perry}
\begin{equation}\label{47}
S_{ij}=-\frac{1}{8\pi}(\kappa_{ij}-\delta_{ij}\kappa_{\delta\delta})
\end{equation}\\
where $ \kappa_{ij} = K^+_{ij}-K^-_{ij} $ expresses the extrinsic curvature with ``$+$'' and ``$-$'' for the inner and the outer regions respectively and $ K_{ij} $ gives the discontinuity in the extrinsic curvatures in the second fundamental forms for both shell sides such as,
\begin{equation}\label{48}
K_{ij}^{\pm}=-n_{\eta}^{\pm}\left[\frac{\partial^{2}x_{\eta}}{\partial
\phi^{i}\partial\phi^{j}}+\Gamma_{\alpha\beta}^{\eta}\frac{\partial
x^{\alpha}}{\partial \phi^{i}}\frac{\partial x^{\beta}}{\partial
\phi^{j}} \right]|_\Sigma 
\end{equation} 
where  $ \phi^{i} $ = the intrinsic parameters on the shell's surface, $ n_{\eta}^{\pm} $ indicates the two sided unit - normal to the surface $ \Sigma $ compatible with the metric , 
\begin{equation}\label{49}
ds^2=f(r)dt^2-\frac{dr^2}{f(r)}-r^2(d\theta^2+sin^2\theta\,d\phi^2)
\end{equation}
where 
\begin{equation}\label{50}
n_{\eta}^{\pm}=\pm\left|g^{\alpha\beta}\frac{\partial f}{\partial
x^{\alpha}}\frac{\partial f}{\partial x^{\beta}}
\right|^{-\frac{1}{2}}\frac{\partial f}{\partial x^{\eta}},
\end{equation}
with $ n^{\mu}n_{\mu}=1 $.
In accordance with the Lanczos equation , the surface energy tensor  { $S_{ij}=diag  [{\sigma, -\Xi, -\Xi,
-\Xi }$]
where $ \sigma $ denotes the surface energy density and $ \Xi $ the surface pressure .\\
We get the surface energy density $ (\sigma) $ as
\begin{equation}\label{51}
\sigma=-\frac{1}{4\pi D}\left[\sqrt{f}\right]^+_-
\end{equation} 
which implies 
\begin{equation}\label{52}
\sigma=-\frac{1}{4\pi D}\left[
\sqrt{1-\frac{2M}{D}}-\sqrt{1-\frac{4(2\pi-\alpha\rho_k)\rho_kD^2}{3}}\right]
\end{equation}
and we get the surface pressure as 
\begin{equation}\label{53}
\Xi = -\frac{\sigma}{2}+\frac{1}{16\pi}\left[\frac{f^{'}}{\sqrt{f}}\right]^+_-.
\end{equation}
which implies 
\begin{equation} \label{54}
\Xi=\frac{1}{8\pi
D}\left[\sqrt{1-\frac{2M}{D}} -\sqrt{1-\frac{4(2\pi-\alpha\rho_k)\rho_kD^{2}}{3}} \right]-\frac{1}{16\pi}\left[\frac{\frac{2M}{D^2}}{\sqrt{1-\frac{2M}{D}}} + 
\frac{\left\lbrace \frac{8(2\pi-\alpha\rho_k)\rho_kD}{3}\right\rbrace
} {\sqrt{1-\frac{4(2\pi-\alpha\rho_k)\rho_kD^2}{3}}}\right]
\end{equation} .\\
The Equation of State parameter $ w $ is given by 
\begin{equation} \label{55}
w = \frac{\Xi}{\sigma} 
\end{equation}
Substituting the above parameters, we get 
\begin{equation}\label{56}
w(D) = -\frac{1}{2} - \frac{D}{4}\frac{\left[\frac{\frac{2M}{D^2}}{\sqrt{1-\frac{2M}{D}}} + 
\frac{\left\lbrace \frac{8(2\pi-\alpha\rho_k)\rho_kD}{3}\right\rbrace
} {\sqrt{1-\frac{4(2\pi-\alpha\rho_k)\rho_kD^2}{3}}}\right]}{\left[
\sqrt{1-\frac{2M}{D}}-\sqrt{1-\frac{4(2\pi-\alpha\rho_k)\rho_kD^2}{3}}\right]}
\end{equation}\\
For real solutions, we have $ \frac{2M}{D}<1 $ and as well as $ \frac{4(2\pi-\alpha\rho_k)\rho_kD^2}{3} \ll1 $. Expanding the square root terms and neglecting the higher order terms, we get 
\begin{equation}\label{57}
w(D)\approx
-\frac{1}{2}+ \frac{9D}{4\rho_kD^3(2\pi-\alpha\rho_k)-6M}
\end{equation}. \\
Now, we see the mass of the shell which has been given by ,
\begin{equation}\label{58}
m_s=4\pi D^2\sigma
\end{equation}
which on substitution gives 
\begin{equation}\label{59}
m_s=D\left[\sqrt{1-\frac{4(2\pi-\alpha\rho_k)\rho_kD^2}{3}} - \sqrt{1-\frac{2M}{D}}\right]
\end{equation}.\\
where $ M $ being the Gravastar's overall mass which can be obtained as 
\begin{equation}\label{60}
M = -\frac{4\alpha \rho_k^{2}D^{4} + 2\sqrt{3}Dm_s\sqrt{4\alpha\rho_k^2D^2 - 8\pi\rho_kD^2 + 3} + 8\pi\rho_kD^4 - 3m_s^{2}}{6D}
\end{equation}.\\

\section{Conclusion}\label{sec7}

This paper has analyzed a unique Gravastar scenario in extended $ f(R,L_{m}, T) $ gravity with respect to the original gravitational vacuum star model put forward by \cite{Mazur} \cite{Mottola}. This was proposed as one of the alternatives to the Black Hole to resolve the singularity and the event horizon problem that the Black Hole theory has faced. Gravastar is similar to Black Hole but singularity free and without event horizon. The Gravastar is characterized by three different regions with distinct equations of state : the interior region with the EoS $ p = -\rho $ , the intermediate thin shell with the EoS $ p = \rho $ , and the exterior that is completely vacuum given by Schwarzschild solution with EoS $ p = \rho = 0 $. Analysis has been done on the multiple physical parameters of those regions in the specified model of the $ f(R,L_{m}, T) $ gravity manifests as:\\

\textbf {Interior region}: Utilizing the EoS for the interior space-time, we have obtained the singularity free solution, and also the non-conservation of the energy equation. The Figure \eqref{1} depicts that the Gravastar is being free from singularity in the interior region.\\

\textbf{Intermediate thin shell} : Implementing the EoS $ p = \rho $, we have explored the shell conditions required for the metric potential, which consists of extremely relativistic fluid, and the thin shell plays a vital role to serve as a Black Hole alternative. Hence, detailing of various physical properties in the shell has been done in the following :\\
(1)\textbf{Pressure and matter density}: Pressure and matter density have been found in the shell region of the Gravastar as a function of the radial co-ordinate $(r)$. Figure \eqref{2} demonstrates the behavioural changes with respect to the radial coordinate in the shell.\\
(2)\textbf{Proper length of the shell}: The appropiate shell thickness  has been obtained as a function of $ r $ in this paper,where we have assumed the thickness to be very small.\\
(3)\textbf{Energy content}: We have discovered a connection between the Gravastar's energy density versus the radial co-ordinate $ r $, of how it depends on $ r $ in the thin shell region.\\
(4)\textbf{Entropy}: The entropy of the shell region has been resolved in this paper  assuming the gravitational units $ G = c = 1 $ with the Planckian units $ k_B = \hbar = 1 $ .\\

\textbf{Junction condition and EoS}: In accordance with the junction condition discussed here, the junction should be smooth enough between the interior region and the exterior region of the Gravastar. From the Darmois-Israel formalism, the metric coefficients are continuous at the junction surface, where we have found surface pressure and energy density and have derived the equation of state and have mentioned the condition for the real solution there. Gravastars or the gravitational vacuum condensate stars have not been discovered or observed scientifically, but many evidence in support of and opposing the LIGO gravitational waves discovery are the result of merging Gravastars or Black Holes. Theoretical evidence of Gravastars are the main contents of this paper in the framework of modified $ f(R,L_{m}, T) $ gravity. Thus, we conclude this result is in agreement with the model of the Gravastar as originally proposed by Mazur and Mottola in the modified $ f(R,L_{m}, T) $ gravity with its physical and spatial features.

 \end{document}